\documentclass{article}
\usepackage[utf8]{inputenc}

\title{Application of Blockchain Smart Contracts  in E-commerce and Government}
\author{Kamal Kishor Singh (kamal.kishor@gmail.com)}

\date{24 March 2022}

\begin{document}

\maketitle

\textbf{ABSTRACT}

With technological advances and the establishment of e-commerce models, business challenges have shifted to online platforms. The promise of embedding self-executing and autonomous programs into blockchain technologies has attracted increased interest and its use in niche solutions. The precarious nature of business transactions and contracts has been brought to light with several scandals such as money laundering and illicit transactions, resulting in the loss of billions of moneys globally. The failures to prevent fraud, insecurities, financial crimes, and other illegal transactions have hampered the efforts of free and fair markets. The right solution perceived to foster security, accountability, inclusiveness, cost-effectiveness, and transparency involves smart contracts. Nonetheless, smart contracts are not well-understood due to their immaturity as an emerging technology. This research paper's principal aim is to analyze the implications of blockchain smart contracts on the modern e-commerce business. Using qualitative interviews, the study sought the opinions of the eleven industry leaders regarding smart contracts. Findings reveal that the technology is gaining momentum in e-commerce, particularly in financial transfer, record-keeping, real estate and property management, insurance, mortgage, supply chain management, data storage, authorization of credit, denaturalized intelligence, aviation sector, shipping of products, invoice financing and other domains. The significant benefits of widespread adoption and deployment of smart contracts include their capability to deliver decentralization, efficacy, cost-effectiveness, transparency, speed, autonomy, transparency, privacy, and security, encouraging the emergence of novel business models. Albeit these benefits that revolutionize online transactions, the technology faced multifaceted challenges. Smart technologies are only a decade old and are not advanced in security, transparency, cost-effectiveness, and regulatory framework. Furthermore, organizational, and technical challenges limit their deployment: incompatibility with legacy systems, scalability, bugs, speed, and lack of talent and understanding regarding smart contracts. Consequently, policymakers, developers, researchers, practitioners, and other stakeholders need to invest effort and time to foster the technologies and address pertinent issues to enable the global adoption of smart contracts by small and big businesses.  

\section{Introduction}
\subsection{Background }
The global economy is experiencing rapid technological advances and a digital value chain coupled with process automation. The growing technological systems are encouraging changes in business aspects. One of the significant shifts experienced involves automation or the digitalization of commerce. Tapscott (2016) notes that technology facilitates efficient and cost-effective operations with less human resources and minimizes operational costs. Additionally, it speeds up the different processes and results in a reduction in the overall business cost. Technology can speed up various processes and brings about outstanding safety and transparency in business (Tapscott and Tapscott, 2016). Digitalization entails the utilization of technologies to transform business models and offer novel revenue streams and value-generation opportunities in digital businesses (BarNir, Gallaugher and Auger, 2003). Digitalization offers novel ways of communicating, collaborating, transacting, interacting online, and processing work (Zhang, Bian and Zhu, 2013). Many businesses globally are reinventing and transforming their commercial operations from manual to e-commerce to increase efficiency, security, productivity, cost-effectiveness, and ease of access to information (Daniel and Wilson, 2002; Ismanto et al., 2019). 
One of the technologies that have received widespread attention in the last three years is blockchain technology, there will be transformational impact of Blockchain Technologies on strategy and operating process on Industries which includes Pharma, Mechanical Engineering and Electronics, Healthcare, Insurance tourism, Construction, Public administration. The speculation on blockchain's rising values is rampant and is evident with the first and most popular application of bitcoin that grabbed headlines for its volatility and skyrocketing prices. The blockchain concentration is not surprising, primarily due to the surge of its market value from below USD 20 billion to over USD  200 billion in 2017 (Katsiampa, 2019). (Reference: Deloitte's 2020 Global Blockchain Survey) 
 
However, the blockchain application goes beyond bitcoin as businesses, industries, and governments now embrace it. In the World Economic Forum, bitcoin is an important topic, with research estimating that 10 per cent of the worldwide GDP will be stored in blockchain by 2027 (Beneki et al., 2019). Recently, many governments and analysts have published several reports regarding the potential impacts of blockchain (Blockchain: The India Strategy, Land Registration  
Blood Bank, Excise Services, Blockchains for Government: Use Cases and Challenges, OECD Global Blockchain Policy Forum Report). Noteworthy, large investments are being made on the blockchain, with venture-capital funding start-ups proliferating to USD 1 billion in 2017 alone (Katsiampa, 2019). Additionally, the blockchain-based investment model of initial coin offerings (ICOs) involving the sale of cryptocurrency tokens is rapidly growing. Leading technological players are also rapidly investing in the technology, with IBM employing over one thousand staff and investing more than USD 200 million in blockchain technology (Carson et al., 2018).  
The limited contact between sellers and customers, personal data misuse, escalating fraudulent activities, and commission fees force businesses to seek better and innovative approaches to conducting business processes (Ismanto et al., 2019).  
Canada, Mexico, USA, and many more countries are using blockchain for various applications. 
Across global financial services, government supply chains, healthcare, businesses, and other industries, innovators are examining the feasibility of deploying blockchain to reinvent and transform conventional business models.  
(References: Which Governments Are Using Blockchain Right Now? | ConsenSys, Blockchains for Government: Use Cases and Challenges, OECD Global Blockchain Policy Forum Report) 

Therefore, blockchain technology has already generated substantial interest in various industries. As the areas of applications increase for blockchain, the leaders customize the technology to fit their needs (Ream et al., 2016). A blockchain is described as a decentralized network technology created to facilitate bitcoin cryptocurrency. Interests in technology have risen from its inception in 2008 and escalated with smart contracts' reinvention (Singh et al., 2020). The blockchain-center smart contracts involve self-executing codes on the blockchain that automatically implement agreements between actors as a crucial step in streamlining processes presently spread in various databases and ERP systems (Ream et al., 2016). The promise of embedding self-executing and autonomous programs into blockchain technologies has attracted increased interest and its use in niche solutions (Ismanto et al., 2019). The primary reasons for increased interest in the technology involve its capability to deliver decentralization, enhance security and integrity, and enable the emergence of novel business models. 
The industry leaders are already utilizing blockchain and have achieved outstanding business benefits, including enhanced security, improved efficiency, and greater transparency, speed of 
transactions, enhanced traceability, and cost-effectiveness. Smart contracts deploy blockchain technology to offer computer protocols that digitally simplify, enforce, and authenticate contracts' negotiation and performance. Smart contracts foster credible transactions while eliminating third parties' roles (Singh et al., 2020). Due to the decentralization experienced in blockchain among all permitted actors, it does not require go-betweens or intermediaries. As a result, it minimizes conflicts, saves time, and offers cost-effective, rapid, and more secure options than conventional systems. Consequently, many businesses, governments, and businesses are embracing smart contracts. 
Indeed, several business and technical analysts have claimed that blockchain will or is revolutionizing business and redefining economies. While such enthusiasm is shared due to its potentiality, businesses should also worry about the hype. According to Iansiti and Lakhanim (2017), blockchain not only introduces security problems but faces a myriad of technological, organizational, societal, and governance barriers. It could be imprudent to rush into deploying the technology without extensively understanding the strategies to use to take hold of the technology. While its impacts and benefits are enormous, it might take decades for the technology to seep into the global economy, businesses, and social infrastructures. The process of its deployment will be steady, but not instantaneous or sudden, as the wave of technological innovations and organizational change takes momentum (Iansiti and Lakhani, 2017). The benefits and strategic implications of blockchain contract are extensively examined in this exploration by using case studies of companies currently deploying the transformative technology. 
 
\subsection{Business Problem Statement  }
The precarious nature of business transactions and contracts has been brought to light with several scandals such as money laundering and illicit transactions, resulting in the loss of billions of moneys globally (Notland et al., 2020). According to Edelman (2017), the business world is experiencing a “trust crisis.” The recent decades have seen debate revolving around discovering and implementing ideal mechanisms intended to foster a free-market economy with robust security, accountability, inclusiveness, and transparency. The failures to prevent fraud, insecurities, financial crimes, and other illegal transactions have hampered the efforts of free and fair markets. With technological advances and the establishment of e-commerce models, business challenges have shifted to online platforms (Notland et al., 2020). A major solution perceived to 
foster security, accountability, inclusiveness, cost-effectiveness, and transparency involves smart contracts. Smart contracts in the commercial field are not yet proven, but various industries believe that permissioned blockchains maintained by a small group of actors will soon find significant adoption (Ream et al., 2016).  
Despite the hype about blockchain technology, it is still perceived as an immature technology, and the market is still emerging. Therefore, a straightforward recipe for its success is yet to emerge (Carson et al., 2018). Unstructured experimentation on blockchain smart contract solutions without a precise strategic examination of its value and its feasibility in capturing its values by businesses will limit companies' ability to realise a return on their investment (Carson et al., 2018). With that in mind, businesses need to determine whether blockchain technology has strategic values that justify key investments. This research intends to answer such questions by investigating smart contracts' strategic importance and the value that businesses can capture, including the approaches to utilise the technology.  Industry and business analysis using primary data delivers a detailed analysis of the e-commerce industry, including the value chains, to determine smart contracts' applicability as a solution to emerging challenges. Additionally, these concepts have been characterised by a comprehensive review of literature.  
\subsection{Research Aims  }
This research paper aims to analyze the implications of blockchain smart contracts on the modern e-commerce business. The following objectives guide the study: 
	• To examine the applicability of blockchain technology in e-commerce. 
	• Examine the benefits of blockchain and smart contract in e-commerce  
	• Analyze the challenges that limit businesses from applying blockchain technologies and offer a recommendation to enhance the technology usage. 

\subsection{Research Questions  }
I	In which areas of e-commerce can blockchain technology and smart contract technology be applied? 
II	What are the beneficial impacts of deploying blockchain and smart contracts in e-commerce? 
III	How will the blockchain and smart contract impact the future of e-commerce businesses?   
IV	What challenges limit the deployment of these technologies in business and what are the recommendations to overcome these challenges? 

\subsection{Significance of the Study }
E-commerce transactions can be costly due to inefficiencies in the widely utilised legacy systems. Many transactions rest in several intermediaries like dealers, banking services, brokers, and agents, among others. Smart contracts are not widely utilised in e-commerce since the concept is not well-understood due to their complexity and novelty. This exploration enhances understanding of smart contracts, particularly the applicability in e-commerce business models. Consequently, it offers practitioners, businesses, and individual’s much-needed knowledge to foster widespread adoption to minimise transactional costs, chaotic business processes, and inefficiencies associated with traditional transactions. As a result, businesses will adopt the more efficient smart contracts as they discard the inefficient conventional approaches and save millions of funds lost due to costly procedural and intermediary-dependent transactions. 
Additionally, the research informs policy-makers and regulators regarding the right mechanisms that can be developed or improved to tackle the challenges associated with smart contracts and the more extensive blockchain technologies. Ultimately, the study intends to achieve a robust understanding of a better environment for smart contracts deployment. Besides, the study adds to the pool of literature on blockchain and specifically smart contracts that can be insightful to other scholars and learners intending to understand the subject.  
\subsection{Outline}
  
This paper is structured into various sections. Chapter 2 presents literature on blockchain technology and smart contracts and focuses on smart contract advantages and challenges. Chapter 3 covers the methodology utilized to achieve the study's objectives, including procedures and design for data collection and analysis. Chapter 4 details the findings obtained from the respondents. Chapter 5 includes a discussion and conclusion section deliberating on findings and deductions from the study and recommendations for improving smart contracts.

\section{LITERATURE REVIEW }
To find high-quality and relevant studies addressing smart contracts from online databases, a search was conducted. The search terms include smart contract, blockchain technology in business, smart contract, and blockchain technology. Based on these keywords and phrases, the researcher used Boolean operators: ‘OR’ and ‘AND’ to combine the keywords: ‘smart contracts’ OR ‘blockchain technology’ and ‘smart contract’ AND ‘blockchain’. The databases searched include ERIC, ScienceDirect, Business Source Premier, Social Science Citation Index, Google Scholar, Scopus, and IEEE Xplore. The databases yielded relevant studies regarding smart contracts and their use in business.  
\subsection{Blockchain Technology  }
Blockchain technology was initially created as an accounting approach for virtual currency bitcoin (Yaga et al., 2019)). The technology deploys distributed ledger technology (DLT) and is utilized in various commercial applications. According to Lipton (2018), blockchain is majorly deployed in verifying transactions, especially in digital currencies, but it possible to digitize various codes and subsequently attach them in any document within blockchain. The process develops an indelible record that cannot be altered. Moreover, such records' legitimacy can be verified by a whole community deploying blockchain rather than a single centralized system (Ryan, 2017). The high level of security exhibited by public cryptocurrencies has made the world embrace blockchain innovations in fostering efficacies and other intangible technological returns (Yaga et al., 2019). 
A blockchain comprises a block, which involves the current segment recording some or the whole recent transactions. Every time a block is completed, a novel one is produced. According to Pan et al. (2020), once a block is completed, recorded in a permanent database, and numerous blocks within a blockchain are linked to one another (or chained) in a chronological and linear sequence. A blockchain comprises a hash for the previous block and complete information regarding various users’ addresses and their balances (Yaga et al., 2019).  
A blockchain transaction goes through initiation, authentication, block creation, validation of the block, and block chaining. The initiation process involves when a single party (the sender) creates a transaction and conveys it to a network (Yaga et al., 2019). Subsequently, the transaction is authenticated through the nodes (users or computers) of the peer network receiving a message and authenticating the validity by decrypting its digital signature. The next phase involves creating 
a block, whereby the pending transactions are assembled in a ledger known as a block by a node in the network (Ali et al., 2020). Subsequently, the validator node of the network will receive the block and undertake the validation process based on an iterative process which calls for consensus from the majority network (Ryan, 2017). After the transactions have been validated, a novel block will be chained into the blockchain and will be broadcasted to the network.  
Blockchain consists of sequences of blocks that comprise a complete list of all transactional records similar to a traditional ledger (Pan et al., 2020). Every block points to the previous block using a reference known as a hash value. The previous block is known as the parent block, while the first block within a blockchain is labeled as the genesis block.  
  The maximum amount of transactions that can be stored on a block relies on the block size and every transaction's enormousness. Aitzhan and Svetinovic (2018) indicate that blockchain technology deploys asymmetric cryptography mechanisms for validating the legitimacy of transactions. 
As noted earlier, before the documentation of a blockchain transaction in a distributed ledger, the transaction must be validated through consensus whereby the nodes agree if a given transaction and block are legitimate or not (Pan et al., 2020). Nodes include the parties that can vote on the legitimacy of a transaction. It entails a process of agreeing on some data values required for computation. In a distributed and decentralised environment, reaching a consensus can be challenging due to various entities acting based on their interests and the information delivered to them (Tönnissen and Teuteberg, 2020). After a transaction has been broadcasted to the networks, the existing nodes can include the same transaction as a copy in their ledger or ignore the transaction. If most of the entities within a network decide a single state are acceptable, they accomplish a consensus. There are various approaches utilised to reach a consensus: Proof of Work (PoW), Peer-to-peer consensus, Kraft, and Proof of Stake (PoS) (Ali et al., 2020).  
Proof of work involves a protocol to prevent cyber-attacks like distributed denial-of-service (DDoS) that can exhaust the resources within a network or computer by sending numerous fake requests. According to Kouhizadeh et al. (2020), PoW requires that expensive computer calculations are defined also known as mining that should be undertaken to create a novel group of trustless transactions within a blockchain ledger. The mining process usually validates the legitimacy of transactions or evades double-spending and creates a novel digital currency by rewarding users performing mining for completion of a given task. Transactions verified through PoW are usually stored in a public blockchain. Even though PoW minimise frauds, it consumes a significant amount of power, particularly for large networks (Kouhizadeh et al., 2020).  
For PoS mechanism, it comes as an alternative to PoW intended to save power. PoS mi9ners need to prove the ownership of the amount of currency, and it is assumed that people with massive ownership currency are less prone to attack a network. Pan et al. (2020) indicate that the selection using users’ account balance unfair since the single wealthiest individuals take a dominant position in the network. Consequently, most solutions in PoS are proposed with a combination of the stake size to establish the one that will create the next block.  
Another validation mechanism is called Kraft and was put forward by Zheng (2016). Kraft the method ensures that a block is produced at a reasonably stable speed since high speed can result in compromised security. Thus, Kraft utilises the Greed Heaviest-Observed Sub-Tree (GHOST) rule to resolve the security challenges (Cole et al., 2019). Rather than using the longest schemes, the GHOST approach will weigh the branches, and miners can choose the best follow. To achieve non-interactive proofs to retrieve the past state snapshots, Chepurnoy and Saxena (2016) advanced the Peer-to-peer consensus algorithm. The protocol enables the miners to store the past block headers rather than the entire block.  
\subsection{Smart Contracts  }
A legal scholar and a cryptographer, Nick Szabo, found in 1994 that decentralised ledger could be deployed in smart contracts (also known as self-executing contracts, digital or blockchain contracts) (Szabo, 1997). Szabo believed that contracts would be transformed into computer programs or codes then stored and replicated on systems as well as supervised by computer networks than run blockchain. As a result, ledger feedback could be obtained, like when money is transferred, or products are received. Smart contracts' emerging role is to assist businesses, institutions, and individuals to exchange shares, lease of assets, possessions, real estate, property, 
or other things perceived to have value in a transparent without conflicts. It ensures that the role of the middleman is avoided. Fries and Paal (2019) define smart contracts as computer programs or transaction protocols intended to automate the execution, control, and documentation of legally relevant undertakings and events in harmony with contractual agreements and terms. Further, Savelyev (2017) note that smart contracts' goals involve reducing the need for trusted intermediates, enforcement costs, and arbitrations while reducing or curbing fraudulent losses and accidental exceptions.  
Smart contracts are not well understood, especially among legal practitioners and researchers. This is because these contracts are considered intelligent and can be achieved automatically (Suliman et al., 2018). Nevertheless, even though smart contracts are achieved automatically, every member must accomplish their responsibility. The significant variations between conventional and smart contracts are the manner they are written, how parties agree on the terms, and their legal implications. These unique attributes provide the benefits and drawbacks of the two contracts. However, self-executing contracts have a long history, such as the “on-demand guarantees” (Salmerón-Manzano and Manzano-Agugliaro, 2019). Even though the on-demand does not follow any specific or general concepts of a contract, it is an increasingly specialised institutional context whereby it is probable to codify transactions to achieve self-executing right. Besides, Smart contracts have a particular and narrow user context whereby multiplicities implies within a contract can be regulated (Suliman et al., 2018).  
There are various areas in which smart contracts can be applied: financial transactions, the Internet of Things (IoT), and prediction markets. In financial transactions, smart contracts are increasingly suitable for equity crowdfunding, online insurance, and peer-to-peer lending. Conventional financial trade requires to be coordinated by a central entity like a central clearing organisation or exchange but the agility of smart contracts minimises transaction costs while enhancing efficiency and avoid cumbersome work (Wang et al., 2018). Also, smart contracts are applicable in prediction markets by offering better future predictions, speculation and direct hedging mechanisms. This can be attributed to the distributed consensus verification as well as immutability. Tw major smart contracts include Gnosis and Augur whereby Augur enables the creation of an accurate prediction tool using blockchain technology (Wang et al., 2018). Internet of things (IoT) is another area where smart contracts have found use where the two are combined 
to foster information sharing between devices and enable individuals to automate time-consuming processes in a cryptographically veritable way.  
\subsection{Smart Contracts in E-Commerce  }
Incorporating smart contracts into electronic commerce (e-commerce) transforms the shopping experience for both businesses and consumers. It delivers value to the vendors, consumers, and sellers by fostering trust between these parties and streamlines their processes (Udokwu et al., 2018). E-commerce offers perfect platforms for developing smart contracts that can interact with tokens. The token ID is linked to the products that consumers purchase, and the consumers can interact with their various tokens independently. Conoscenti et al. (2016) argue that in e-commerce, smart contract development ensures that blockchain contracts can be executed only when the users meet specified requirements of a given deal. For instance, when the consumers check out of an e-commerce site, the payment they make can be kept in smart contracts. Subsequently, the merchant advances to authorise a change of ownership of the products and link the contract to the shipping entity. (Udokwu et al., 2018) After these requirements are fulfilled and acknowledged, a smart contract can release the money to the sellers.  
Smart contracts foster electronic payments using cryptographic proof, which hashes and provides the timestamps of various transactions on a chain. Thus, various buyers and sellers can transact directly while eliminating a trusted third party (Ryan, 2017). Thus, it streamlines e-commerce and minimises mistakes, tax evasion, corruption, and fraud as a reliable online tracking system. With e-commerce exchanges, there is minimal risk that one party might not fulfill her deal since a smart contract executes the payment. This saves time and cost for the sellers and buyers as they actively fulfill their obligations (Udokwu et al., 2018). The exchanges achieve immutability and immediacy, and they feel like the consumer is using cash.  
Smart contracts foster reciprocated trust between the buyers and sellers to minimise the exposure of sensitive information to a third-party application that users do not have control over (Conoscenti et al., 2016). Smart contracts also offer a transparent and consistent approach of enabling commodities to cross borders since all the entities along the supply chain have access to a copy of the same ledger that exhibits the origins and shipping of products while traveling from the sellers to the buyers (Tanja, 2018). This considerably minimises errors of unreliable record-keeping. Consumers can track and check the products they have orders to ensure the order is in the 
right phase or place during the delivery process. Distributed ledger transactions can be stored publicly to enable consumers to easily access and control their order delivery (Ryan, 2017).  
Several online or e-commerce businesses deploy smart contracts, and companies developing and delivering e-commerce solutions to other businesses. For instance, Elysian entails a platform that offers assistance for businesses intending to develop e-commerce company websites using blockchain smart contracts (Tanja, 2018). Elysian intends to develop and offer revolutionary e-commerce ecosystems built on security, transparency, and trust. Besides, Elysian has delivered various solutions, including virtual reality and augmented intelligence, that enhance shopping experiences. Another platform includes E-Commerce, a decentralised market whereby the users are considered the owners of digital assets like brands and stores (Suliman et al., 2018). E-commerce platform enables users to record the ownership of their shop through blockchain smart contracts (Tanja, 2018). Subsequently, the owners can sell or move their digital assets since they have full control over the store and their assets. For consumers, the use of an E-Commerce platform reduces the prices and ensures that constant discounts are available. 
OpenBazaar is another platform based on P2P experiences. The platform is free for use and can be accessed online or even downloaded as an application. OpenBazaar is equated to the conventional Amazon and eBay platforms, but it provides greater control to the users over their information and enables consumers to pay for their products using 50 altcoins. Furthermore, Eligma offers an ideal platform that uses blockchain smart contracts for e-commerce to manage online shopping (Tanja, 2018). Eligma aims to ensure that shoppers experience less complicated processes and are more comfortable while shopping. Eligma deploys artificial intelligence to help consumers save time while monitoring their online shopping and finding the best deals. Besides. The company offers distributed ledger transaction programs where consumers are awarded for shopping via the platform by receiving tokens of ELIs (Tanja, 2018). Such tokens can be utilised to undertake additional purchases, which encourages the use of blockchain smart contracts in e-commerce.  
\subsection{Advantages of Smart Contracts in Business }
The deployment of business can be attributed to far-fetched benefits of blockchain, including greater transparency, increased speed of transaction and efficiency, as well as enhanced security, reduction of business costs, and improved traceability (Tapscott and Tapscott, 2016). Transactional histories are becoming increasingly transparent due to the deployment of blockchain technologies. 
As a distributed ledger, the actors of the network will share similar documentation instead of individual copies. To alter one transaction, it needs to change all subsequent documents and to collude with the whole network (Savelyev, 2017). Hence, data on a blockchain has been more accurate, precise, transparent, and consistent than on paper-based imprecise processes.  

\textbf{2.4.1	Security}

Blockchain enhances security through various mechanisms as compared to record-keeping systems. Transactions need to be agreed upon prior to being recorded. Besides, information is usually stored across several computer networks rather than one server, making it challenging for hackers to compromise transactional data (Singh et al., 2020). In any business, safeguarding sensitive data is essential and blockchain technology offers an opportunity to transform the manner in which critical information is handled, which helps to curb fraud and unauthorized activities. Additionally, the miners can benefit from the anonymity that cannot be enjoyed in conventional contracts. For instance, trusted intermediaries facilitating most common sale agreements like credit card organisations need proof of identity to carry out future payments (Conoscenti et al., 2016). Consequently, intermediaries store massive amounts of consumers’ data sensitive and personal to every consumer who utilises such services. This escalates the risk of the consumer’s data being hacked and exploited (Udokwu et al., 2018). Transactions deploying cryptocurrencies enable consumers to buy commodities online without providing personal data. Also, the use of distributed ledger transactions in e-commerce ensures that consumer data will not be stored in a centralised server, reducing the chance of hacking or stealing data (Golosova and Romanovs, 2018). 

\textbf{2.4.2	Traceability and Transparency  
}

Moreover, blockchain improves traceability, particularly for businesses dealing with multifaceted supply chains where it is challenging to trace items back to their origins. The documentation of goods on the blockchain provides an audit trail that depicts assets' origin (Tapscott and Tapscott, 2016). Such transactional data visibility assists in verifying the authenticity of assets and further curbs fraud. Due to the decentralization of data by deploying distributed ledgers, significant transparency is achieved. According to Giancaspro (2017), the absence of central power and intermediary for validation and collation of transactions and the transparency evident in blockchain implies that commercial arrangements undertaken via smart contracts in the public ledger are visible to all miners (Udokwu et al., 2018). As a result, users are likely to trust one another due to increased confidence.  

\textbf{2.4.3	Speed and efficiency  
}

Also, blockchain increases the speed and efficiency of transactions, unlike conventional paper-based processes. It streamlines and automates processes prone to human errors or requires third-party mediations, ensuring transactions to be completed rapidly and efficiently (Savelyev, 2017). Transactions enabled by smart contracts are usually not validated by a trusted intermediary but instead by consensus mechanisms. Therefore, instead of a credit provider, a financial institution, or an insurance organisation facilitating digital transfers of properties, smart contracts will automatically do all these activities (Giancaspro, 2017). Miners in blockchain are considered as contract parties who only need to determine the content of their agreements, and smart contracts will execute the terms automatically. The disintermediation process enhances efficiency by enabling the blockchain to address critical facets of transactions ranging from record-keeping, enforcement, auditing, and monitoring (Wang, 2016). As a result, settlements will occur quickly due to eliminating delays caused by intermediaries in conventional contracts.  Further advancement of distributed ledger technology will result in instantaneous settlements for multifaceted transactions. Additionally, automation of critical processes in the lifecycle of a contract can minimise human interventions, which further improves efficiency (Giancaspro, 2017).  

\textbf{2.4.4	Cost-Effectiveness  
}

Furthermore, blockchain reduces business costs since many third-parties are eliminated from being middlemen and reducing legal and transaction costs. The elimination of a central authority and trusted intermediary coupled with how transactions are transparently verified and subsequently added onto the chin by the miners implies that most transactions and legal costs incurred due to intermediation are eliminated (Giancaspro, 2017). These costs include administration, legal and service fees linked to preparing, executing, and monitoring written contracts. For instance, when a contract is established through credit card purchase, the consumers purchase items from merchants and pay through credit cards, the merchants will apply surcharges. Accordingly, surcharges involve the cost of the merchant for accepting payments through credit cards. Moreover, the credit card organisations apply additional fees. All these costs can be eliminated by deploying smart contracts (Giancaspro, 2017). The cost-savings associated with deploying smart contracts surpasses the transactions to minimise infrastructure costs (Walch, 2015). Essentially, avoiding an intermediary by deploying smart contracts enables consumers, businesses, and 
governments to minimize commercial and operational costs drastically. Consequently, substantial overheads are reduced (Giancaspro, 2017).  

\textbf{2.4.5	Accurate Recording-keeping  
}

A major requirement of smart contracts involves the terms and conditions being defined explicitly and precisely. Such requirements are essential because any omission can result in transaction errors. Thus, automation of contracts evades the challenges of manual filling of forms (Golosova and Romanovs, 2018). Also, since transactional information is stored in a distributed ledger, including several computers or nodes, it is hard to alter anything. Therefore, an e-commerce that deploys smart contracts ensures all information and documents are kept in their original versions with several copies, making data loss improbable (Conoscenti et al., 2016).  
\subsection{Challenges of Blockchain and Smart Contracts 
}
\textbf{2.5.1	Image challenge } 

There is a common misconception among many users that blockchain technology only involves Bitcoin applications and cryptocurrencies. Particularly, crypto has a negative reputation associated with the hackers and fraudsters that deploy the technology for criminal undertakings (Meijer, 2020). This is a negative reputation for blockchain applications as a whole, and many businesses and individuals who do not understand that crypto is just a single application of blockchain might shy away from adopting it for smart contracts. Therefore, before the widespread adoption of blockchain smart contracts is achieved, the public needs to comprehend the differences between bitcoin and other cryptocurrencies and blockchain (Conoscenti et al., 2016). Such understanding will assist in eliminating negative reputation and can lead to widespread willingness to deploy smart contracts. Meanwhile, several collaborative initiatives are being established in blockchain domain intended to realise a wider change. Such interdependence might assist in moving ahead. Nonetheless, some businesses fear that blockchain's disruptive nature, while others do not perceive blockchain contracts as disruptive (Meijer, 2020). Some fear that they might lose their market share overnight when blockchain contracts become the order of the day due to being obsolete.  

\textbf{2.5.2	Regulations and vested interests  }

The current regulations represent substantial challenges for blockchain technologies since the prevailing regulations existing companies and individual developers over the disruptors (Conoscenti et al., 2016). The digitalisation era is occurring in a heavily regulated environment 
due to the long-standing regulations enacted by various regimes intended to safeguard consumers and property rights (Meijer, 2020). Blockchain presents novel hurdles to regulators who intend to safeguard the markets and consumers, and the inflexibility with which developed economies are approaching blockchain technology might serve to strangle its growth and innovations. However, such perspectives are rapidly changing as regimes and businesses are experiencing benefits from the technology and are establishing a friendly regulatory environment that fosters innovations while safeguarding the consumers.  

\textbf{2.5.3	Security and Privacy }

Blockchain technologies such as cryptocurrencies deliver pseudonymity, and other potential applications like smart contracts and transactions call for identities to be linked indisputably (Meijer, 2020). As a result, they raise significant questions concerning the security and privacy of data accessible and stored on a shared ledger. Many businesses are using privacy rules based on regulations governing business conduct. Consumers entrust businesses with sensitive data, and when such data is stored within a public ledger, it is unlikely to be secured and remain private. Only limited scenarios in smart contracts provide good protocols to cope with security challenges (Conoscenti et al., 2016). Even though blockchain technologies are more secure than conventional systems, sophisticated hackers are still breaching businesses, applications, and systems deploying blockchain technologies. For instance, the Ethereum blockchain was attacked recently in what is labeled as Classic 51
Failed smart contracts have been examined and categorised into prodigal, greedy, and suicide contracts (Nikolić et al., 2018). Prodigal contracts entail those hacked, and such fraud has led to cryptocurrencies becoming properties of fraudsters (Giancaspro, 2017). Suicide contracts involve the ones that have been closed after the exit requirements have been triggered by an individual undertaking an attack. This can be caused by a wrongly executed exit clause leading to a wrong individual taking all the encrypted money involved in smart contracts (Abadi et al., 2016). According to Atzei et al. (2017), insufficient protection is to blame for smart contracts allowing money to be transferred illegally. The greedy contracts can arise from bad practices or miswriting and cause the contracting party not to have the legitimacy of receiving encrypted currency. When 
such issues occur, the contracts are likely to end, leading to economic losses attributed to vulnerability failures (Golosova and Romanovs, 2018).  
According to Wang et al. (2018), smart contracts experience re-entrancy vulnerability whereby an intruder can deploy a recursive call function to undertake several repetitive withdrawals, but the attacker's balance is deduced just once. This can cause unexpected behaviours, including consuming all the gas. Also, transaction-ordering dependence (TOD) is a significant challenge in a smart contract whereby several dependent transactions invoking a single contract are situated in the same block (Conoscenti et al., 2016). Miners usually set arbitrary orders between various transactions, including the contract’s final state. Consequently, an intruder can effectively launch an attack when the transactions are not executed in the right order (Wang et al., 2018). 
Additionally, the timestamp dependence is a major loophole where the miner can alter the value while other miners accept the block. A security challenge that arises by using timestamps as the activation mechanism to undertake specific activities includes easily manipulated contracts. Furthermore, smart contracts lack reliable data feeds and privacy issues since all history information is stored in a blockchain and can be seen by other users (Wang et al., 2018).  
The solutions to security and privacy is not only through the development of policies by the governments to safeguard privacy and security. Self-sovereign identities within the blockchain can allow businesses to capture and regulate their data (Meijer, 2020). While several protocols have been proposed like proof of zero-knowledge and good identity initiatives are being developed like Sovrin, businesses will still have to wait for a radically novel identity framework.  

\textbf{2.5.4	Lack of clear governance and regulatory framework }

There is inadequate clarity regarding the regulatory environment behind blockchain technologies, which affects mass adoption in business. Regulations seem to be struggling in keeping up with technological advances, particularly the hyped blockchain. A major motive of blockchain contracts is to minimise oversight, but this remains one of the major hurdles today (Meijer, 2020). Most organisations are developing the technology as a means of transaction, but there are no specific smart contracts regulations. Therefore, no business adheres to specific regulations when deploying smart contracts. Since regulations do not cover smart contracts, they have inhibited its widespread adoption and investments.  
Notwithstanding the bottlenecks and challenges, centralised systems, especially within the banking and financial industry, can act as shock absorbers during a crisis (Golosova and Romanovs, 2018). Conversely, decentralised systems might be less dependable during a crisis, affecting the users directly unless adequate measures are put into place. It has been proposed that smart contracts need to work with existing regulatory structures. From a legal perspective, smart contracts need to tackle various issues like jurisdictional, trial and enforcement risks (Salmerón-Manzano  and  Manzano-Agugliaro, 2019).  

\textbf{2.5.5	Maturity and Scalability Problems  }

Blockchain several implementation hurdles since it is a relatively young tech. The technology is less than a decade old and needs time to mature. In various consortiums, the players solve decentralized puzzles and come up with unique solutions such as Ripple, Corda, Enterprise Ethereum, and Hyerledger (Golosova  and  Romanovs, 2018). Due to maturity challenges blockchain technology faces the scalability challenge of its network that limits the adoption, particularly public blockchain. Legacy systems and networks are known for their capability to process thousands of transactions every second. For example, Anupam (2019) compares the largest centralised payment system, VISA, and Bitcoin's biggest crypto payments. Where Visa processes 65,000 transactions per second, the maximum speed for Bitcoin is 7 transactions each second. For centralised architecture, the controlling authority will decide the flow and does not always notify its peers regarding a given transaction, which saves speed and time (Anupam, 2019). Conversely, for blockchain technology, the validation can take several minutes since most nodes need to authorise a transaction. For Ethereum, it can only handle bout 20 transactions every second.  
Scalability challenges is not a major challenges for private blockchain networks because nodes within a networks are specifically developed to process transactions within an environment involving trusted parties that makes a lot of sense business-wise (Meijer, 2020). There are outstanding solutions being developed intended to resolve the scalability challenges like Lightning Network that comprise of additional second layer to the major blockchain network with the intent of fostering rapid transactions. Another solution involves Sharding which groups subnets of nodes to smaller networks known as “shards.” The small network node are responsible for transactions specific to that shard (Meijer, 2020). When deliver together with proof-of-stake consensus mechanism, it has the ability of scaling up.  

\textbf{2.5.6	Limited interoperability and standardisation  }

Besides, blockchain faces a challenges of limited interoperability between massive numbers of blockchain networks. More than 6,500 projects are deploying various, but mostly standalone blockchain platforms as well as solutions with various privacy measures, protocols, consensus mechanism and coding languages. With many different networks, blockchain space faces a “state of disarray” since it lacks globally defined standards that enable networks to communicate with one another (Meijer, 2020). Due to lack of uniformity across the blockchain protocols, consistency is lacking on fundamental processes such as security and this makes its adoption impossible for some players (Golosova  and  Romanovs, 2018). The development of industry-wide standards for various protocols can assist many businesses to cooperate on developing applications, sharing blockchain solutions and validating concepts, and this will make it easier to incorporate blockchain with other existing systems.  
Various projects are underway with the intent of realising interoperability for various blockchain networks like Ark project that deploys SmartBridges architecture to address interoperability problem. The Ark project is considered to deliver global interoperability and enables cross-blockchain communications and transfers. Another project entails Cosmos and it deploys Interblockchain Communication (IBC) protocols to ensure that blockchain economies can operates outside the silos and files can be transferred between networks (Meijer, 2020).  
There a problem of integrating blockchain with legacy systems. In many cases, businesses that decide to deploy blockchain technology are compelled to entirely restructure their existing systems or design them in manner that can successfully integrate the two technologies side-by-side (Golosova  and  Romanovs, 2018). However, many businesses lack skilled developers since they cannot access the required talent to undertake such processes. The dependency on external players to undertake these process force many businesses to invest a significant amount of resources and time to achieve the transition from legacy to blockchain systems. Additionally, there are many incidences of data loss and breaches that have been reported. Such incidences discourage many businesses from shifting to blockchain (Golosova  and  Romanovs, 2018). Every business is reserved or unwilling to alter its databases since data corruption or loss is considered a major risk. Recently, novel solutions are emerging that enable legacy systems to be linked to blockchain backend. For instance, Modex Blockchain Database is designed to assist businesses or users without a 
knowledge in blockchain technology to access its benefits and eliminate the risks associated with loss of sensitive data.  

\textbf{2.5.7	Brain-Drain and lack of expertise }

With the ever-increasing demand for blockchain personnel, the blockchain industry is suffering from lack of sufficient personnel who are skilled and well-trained to develop and manage the multifaceted tasks. The technology demands increasingly qualified and skilled personnel. Meijer (2020) argues that the demand for blockchain-related staff has skyrocketed by about 2000 percent from 2017 to 2020. Therefore, having adequate personnel with the right qualifications is a top challenge in the industry. As noted early, blockchain technology is still immature and evolving. Consequently, it needs time for the developer community to share and for learning institutions, they need to put in place course related to blockchain (Golosova  and  Romanovs, 2018). While such measures might ease market demands, the outcome will become possible only after learners complete their training and this will take time.  
Recent surveys and analyses have demonstrated that a significant number of experts of blockchain are relocating to developed economies while searching for better opportunities. For instance, Indian developers are moving abroad and blame brain drain on the lack of adequate regulatory framework in every nation deploying blockchain technology (Anupam, 2019). Other blockchain developers are reportedly relocating to Switzerland, Singapore and the United Arabs Emirates which deliver tax breaks as well as e-residency for start-ups. The rapidly developing blockchain technological infrastructures in these nations are considered to be better suited for blockchain applications.

\textbf{2.5.8	Complexity and Organisational Challenges  }

Blockchain can be complex, cumbersome and slow attributed to its encryption and the distributed nature. Transactions might take sometimes to process than the conventional payment systems like debit and cash cards (Golosova  and  Romanovs, 2018). When the number of users increase over a network, transactions can even take longer. It can take days to process a single transaction. Consequently, the transactions are likely to cost more than usual and this can limit users from adopting the systems. The challenge extends to blockchain networks that are utilised for other purposes than storing value such as smart contracts and interaction of blockchain and Internet of Things (IoT) (Meijer, 2020). This challenges is likely to be resolved through advanced engineering and processing speeds.  
There are various organisational challenges the e-commerce businesses are likely to incur while deploying blockchain contracts. One of the major challenges involves the lack of awareness regarding smart contracts, especially for small and medium businesses (Golosova  and  Romanovs, 2018). Many businesses do not have sufficient knowledge regarding how smart contracts work due to domination of technicians in the domain and significant technological approach of blockchain. Such problem might hinder many e-commerce businesses from investing and exploring smart contract concepts (Meijer, 2020). A business oriented approach is required to improve the user experiences, particularly for those with limited technical knowledge.  
Also, there is a productivity paradox the efficiency and speed of blockchain networks of executing peer-to-peer transactions come at a cost which is increasingly high for some blockchain applications (Golosova  and  Romanovs, 2018). The inefficiency originates from nodes undertaking the same processes as each node copies data while trying to be the first to offer a solution. Hence, decisions of businesses regarding the adoption and deployment of blockchain smart contracts should be carefully and thoroughly analysed. Returns from individual processing might decline as the networks widen. As a result, blockchain applications need to harness network effects to provide value to the users and the business at large.  
Additionally, there is lack of cooperation. Smart contracts offer the most values for e-commerce when they work together on fields of shared opportunity and challenges. The challenge with most business models is that they are increasingly work as standalone with enterprises developing their own blockchain to beat their rivals (Meijer, 2020). Therefore, in various industries various chains are being created by various enterprises with various standards. Such approach defeats the role of distributed ledge and does not harness network effects making the standalone approach less effective (Golosova  and  Romanovs, 2018). Nevertheless, positive development are being witnessed with the rise of blockchain consortia intend to address industry-wide challenges like critical mass and standards.  

\textbf{2.5.9	Environmental Cost }

Some scholars have warned about the challenges of smart contracts, particularly their resource intensive nature during the verification processes. For instance, Khezr et al. (2019) studied the relationship between bitcoin and sustainability while Tapscott and Tapscott (2017) investigated the association of sustainability and Energy Internet. Energy consumption remains a major hindrance of deploying blockchain technologies. A study conducted by the University of 
Cambridge approximated that the technology consumes high energy than the entire Switzerland (Anupam, 2019). The energy is usually feed into the entire network throughout the lifecycle. Smart contracts follow the bitcoin design and deploy Proof of Proof-of-work (PoW) as a consensus mechanism to validate transactions. Such protocols need the users to resolve multifaceted mathematical puzzles and require massive computing power in verifying and processing transactions and to undertake these processes securely. Meanwhile, the amount of energy utilised by computers competing to resolve a mathematical puzzle is increasingly high. Some approximate that bitcoin transaction consume energy which is almost the annual electricity usage in Denmark (Meijer, 2020).  
To overcome energy requirements, supporters of blockchain technologies are advancing the development of more efficient consensus algorithm that consumes minimal energy. The Proof-of-Stake (PoS) have been proposed as the protocols that combine the user’s stake in the network and the algorithm randomly allocates the process of validation to a given node (Meijer, 2020). Since the users do not need to resolve multifaceted puzzles, the mechanisms can remarkably minimise the amount of energy consumed. Mengelkamp et al. (2018) investigated how laws and private standards can improve sustainability in of smart contracts and depicted how blockchain can enhance sustainability through informing the users regarding the origins of products, offering mechanisms for enforcing representation via smart contract functions of blockchain and ensuring legitimacy. Casino et al. (2019) proposed the use of energy transaction platform centred on peer-to-peer blockchain to foster energy efficiency between prosumers.  Furthermore, de Vries (2019) suggests the creation of an Energy Internet utilizing novel power grid structures of generating renewable energy. In another exploration, Gatteschi et al. (Gatteschi et al., 2018) suggest the deployment of insurance sectors to boost efficiency. Other approaches that can enhance the sustainability of smart contracts include accelerating and automating the exchange of data (Salmerón-Manzano  and  Manzano-Agugliaro, 2019).  
Additionally, storage is a major technical challenge. For instance, Anupam (2019) notes that Bitcoin Core as one of the applications of blockchain requires about 200 GB storage space for every node which is part of the blockchain network. Besides, it needs about 5 GB upload and 500 MB download each day. Therefore, companies struggling with internet speed are disadvantaged in implementing blockchain smart contracts infrastructures (Anupam, 2019).  
\section{METHODOLOGY  }
\subsection{Research Design  }
There are two major research designs: quantitative and qualitative. While some researchers perceive quantitative design to be superior to qualitative research, it is vital to understand that the two do not answer the same questions. Qualitative research focuses on detailed explanation while quantitative design concentrate on numbers or statistics. This study utilised a qualitative research design to answer the research questions. While qualitative studies are perceived to be less rigorous as compared to quantitative studies, qualitative information is usually in-depth and offers breadth regarding the phenomenon under exploration (Creswell, 2014). Also, the qualitative design is beneficial when a phenomenon is multifaceted and cannot be encapsulated by a simple yes or no or any other straightforward answers.  While qualitative data provides numerical data with a clear explanation of the relationship between the variables, qualitative approaches can yield insightful and richer data underlying the patterns manifesting in a phenomenon (Creswell, 2014). Therefore, Qualitative research was found to be the perfect fit for this study due to its novelty.  
The study targeted e-commerce businesses currently deploying smart contracts in their operations, including the leading business leaders in every industry. Purposive sampling was utilized in identifying and choosing information-rich subjects who are well-versed about smart contracts. Purposive sampling is commonly deployed in qualitative research to choose participants, especially those knowledgeable about the interested phenomenon (Creswell, 2007). Besides knowledge and experience, purposeful sampling considers the subjects' availability and their willingness to take part in the study and their capability to communicate their opinions and views in a thoughtful, articulate, and expressive way. The management of these businesses were contacted to assist in recruiting personnel who took part in the study. The researcher utilised purposive sampling to recruit 15 respondents who took part in the study. Heterogonous purposive sampling was used to ensure that the selected sample originates from diverse backgrounds to offer varied experiences and views about smart contracts (Creswell, 2007). Selected subjects received informed consent detailing the purpose of the study. Participation in the exploration was voluntary without any coercion. The privacy, anonymity and confidentiality of the study participants were protected throughout the research.  
\subsection{Data Collection and Analysis  }
There are several data collecting approaches in a qualitative study, including focus groups, observations, content analysis, interviews, and others. The observational approach is also known as a field study and needs a prolonged assessment of a given setting. In content analysis, the researcher examines information fetched from written documents (such as court proceedings) or audio/visual materials (King et al., 2010). The study utilized interview data collection tools to obtain the views, experiences, and opinions of participants concerning smart contracts and their role in business. King et al. (2010) note that interviews are beneficial in gathering people's perspectives and outstanding tool in uncovering the experiences of people to comprehend a given phenomenon. Semi-structured interviews were used whereby the respondents answered to predefined open-ended questions. Semi-structured interviews are based on developed interview guides involving the schematic presentation of the topics or questions that should be explored during the interviewing process (King et al., 2010). Interview guides offer a valuable tool to interview participants about specific aspects systematically. Interviews were conducted via telephone calls that were recorded and subsequently transcribed verbatim in readiness for data analysis.  
For data analysis, there are two general approaches for analysing qualitative data: inductive and deductive approaches (Burnard et al., 2008). In a deductive approach, the researchers uses a structured or predetermined framework for data analysis whereby the investigators impose their theories or structures on the data. Such approach is useful when the researchers understand the likely responses from the participants. On the other hand, inductive approach entails analysing data with minimal use of predetermined theory or structure (Burnard et al., 2008). Inductive analysis is the most common approach particularly thematic analysis of content.  
Thematic analysis was applied to analyse the qualitative data obtained, emphasizing the identification, analysis, and interpretation of patterns of meanings (Adu, 2019). Interview transcripts provided descriptive accounts of the research but do not deliver explanations. Therefore, the investigator need to make sense out of the data by examining and interpreting it. Thematic analysis of the content was improved by the use of NVivo software (Burnard et al., 2008). The first phase of data analysis involved preparing, reviewing and reading the data several times while creating notes about the major ideas emerging from the data. This enabled the researcher to create initial codes which were reviewed and revised and subsequently combined to 
create themes (Adu, 2019). The themes were reviewed to eliminate redundancy and ensure all data is comprehensively described and presented.  
\section{RESULTS  }
This section articulates the views obtained from the participants regarding smart contract. While the study anticipated to recruit 15 respondents to offer opinions regarding smart contracts, only 11 respondents were recruited since other declined to participate in the study due to various reasons like time constraints and unavailability. Therefore, the views presented in this section include views only 11 respondents. The respondents were drawn from e-commerce businesses currently deploying smart contracts in their operations, including the leading business leaders in every industry with sufficient information regarding smart contracts. The subsequent sections present the views of various participants about smart contracts.  
\subsection{The use of Smart Contract E-commerce  }
All the participants were asked whether their organizations utilize smart contracts in their e-commerce undertakings and they agreed that blockchain technologies is an integral part of their business activities. Additionally, the respondents were asked to offered opinions regarding the areas in which they apply smart contracts. The table underneath documents the views of the 11 respondents.  

\textbf{Table 1: Areas of applications of Smart Contracts in E-commerce 
} 
\begin{enumerate}
\item \textbf{Respondent 1 }- Financial industry  	As a financial company, smart contract and blockchain play an integral role in our operations. We use the technology in various transactions including transferring of funds and for the customers to pay for the services or products we are offering them.   
\item \textbf{Respondent 2} - Smart contract is mainly used for record-keeping due to its capability to document. Also, we utilize smart contract to facilitate financial trades, legal processes, credit authorization and crowdfunding agreements or Initial Coin Offering (ICOs). Smart contracts and tokens can be created to foster secure payments.  
\item \textbf{Respondent 3 }- We use smart contracts for storing receipts, customer data, record keeping and managing the general stock.  
\item \textbf{Respondent 4} 	Our real estate business utilizes smart contracts for documenting property ownership from buildings to lands.  
\item \textbf{Respondent 5} 	We use smart contracts to automate payments and rentals of electric vehicle charging stations. Smart contracts enable other parties to add any objects (products or services) to our network  
\item \textbf{Respondent 6 }	We deploy smart contracts in our decentralized threat intelligence. When businesses and ambassadors publish their bounty, they will specify the amount of reward which will be given to any individual submitting the right assertion. When a security expert submits an assertion and it is determined to be accurate, the reward or money promised will be realized automatically to the individual.  
\item \textbf{Respondent 7 	}We use smart contracts in our insurance business for establishing and encoding the terms of policy that will be agreed upon by parties involved. These terms are usually copied into smart contracts and cannot be altered with the involved parties agreeing. Smart contract policy are digitally executed through pre-defined terms. It allows insurance payouts and settlements to be processed promptly and efficiently.  
\item \textbf{Respondent 8 	}We use smart contracts to manage and secure buying and selling of goods on the internet, launch ICO’s and tracking the shipping process as well as stocktaking. Each step in the supply chain can be recorded since smart contracts provide such capability which minimizes theft and missing products. 
\item \textbf{Respondent 9} 	We use smart contracts for flight insurance. It is challenging for our clients to work with airlines to receive compensation for delayed flights, particularly those with travel insurance.  When a passenger’s flight is later for more than two hours, their details are automatically inputted into an application which will digitally notify various compensation options and money will be transferred directly to the client’s credit card. Smart contracts deployed is centered on parametric insurance where the client will not compensation for total loss but instead for losses that occur outside the conventional guidelines of insurance.  
\item \textbf{Respondent 10} 	Our business mainly focus on managing real properties and estate. We use smart contracts to enable secure and efficient buying and selling of real estate.    The owners or brokers can list their properties and buyers can search, view and negotiate on the prices and ultimately buy the properties if their meet their requirements. All the parties involved in a deal can participate in the smart contract and decisive are included to foster fair and legal transactions. A buyer who is interested in buying a property can reserve it by paying about 10 dollars  to escrow firms that holds the property and in case the sellers refuses to change ownership, the buyers can always receive their money back. Paperwork and signatures are undertaken digitally irrespective of the origin of the seller and the buyer.  

\item \textbf {Respondent 11 	}When us smart contracts for invoice financing for obtaining money from unpaid invoices. The invoice purchasers usually must pay upfront in order to takeover an invoice for their business and will get paid the original amount after the debtors have paid the invoice. Therefore, any individual can buy and sell the outstanding invoice through blockchain smart contracts.

\end{enumerate}
\subsection{Benefits of Smart Contracts to E-commerce }
The participants were asked to present the opinions regarding the benefits that are incurred by deploying smart contracts in their respective businesses. Their views are documented in the table underneath.  
\textbf{Views of the Respondent about the Benefits of Smart Contracts in E-Commerce  }
\begin{enumerate}
\item \textbf{Respondent 1  	}Blockchain technology by nature enables autonomy and no outside party will know the owner of the credit cards or obtain their sensitive information. Thus, smart contract foster secure transactions. Once we have recorded a given transaction, it is impossible to edit, delete or alter the transaction. Thus, smart contract are tamperproof and secure in storing important transactions with the fear than someone will access them and alter the number. I believe that while we do not use smart contracts in all our financial undertakings, the areas that have implemented the technologies have realized outstanding security and cases of fraud or corruption have significantly been reduced since workers cannot collude with clients or other parties to alter financial records.  
\item \textbf{Respondents 2 	}The benefits obtained by using smart contracts includes secured encryptions especially for peer-to-peer transactions. In the past, our organization worried about customer’s sensitive data being exposed to external hackers and intruders who can sell information to third party. However, after our organization implemented smart contracts, our systems are more secured and cases of fraud have significantly reduced. Also, our clients are happy since the tiresome middlemen and many parties involved in the transaction have been eliminated. Particularly, today transactions occur at a high speed since when do not need to engage third parties to secure money transferred between the consumers and the company. Also, our esteem clients are assured that they will receive the goods they have purchases since smart contract documentation and record keeping enable them to monitor the shipping process using a unique shipping code. In case of any problem with a given order, our customer representative can ask the client to provide the code which can easily be tracked and the customer can be updated according regarding his/her order. Therefore, smart contracts makes use serve our clients better to their satisfaction. A satisfied client will become loyal and repeatedly purchase from us increasing our long-term financial performance and profitability.   
\item \textbf{Respondent 3} 	A major benefit incurred to our business is cost reduction and efficiency. We continually invest in technologies that can help us save more by streamlines our business activities delivering efficacy. Smart contract automate e-commerce transactions while dealing with another party without the need to engage a third party. As a result, we avoid the cost of third parties. Additionally, the fact that transactions are automated, the processing cost is minimized. Besides, the systems enable our workers to avoid costly human mistakes and errors which can be costly to our business.  
\item \textbf{Respondent 4} 	We had almost trusted in many clients since their records could be altered by our workers or external intruders. However, with the use of smart contract, we have improved the trust in our clients. Cases of altered records or traceable funds are not reported frequently. The technology acts as a backup in case our systems fail or data is tampered with since all transactions are duplicated to enable the client and other parties to have a record of their transactions. Even when moving data to another system or upgrading, we can easily access original documents or records and the probability of suffering from data storage failures are reduced to minimal levels. Also, there is significant cost-savings that stems from automating approval of transactions and clearing of e-commerce computation that were in the past labor-intensive. Smart contracts take up meaningful work that was previously done by accountants and record keepers. There is a reduction in work-hours and a reduction of errors and the time taken to undertake calculations. Since 2019, our company has saved a whopping USD 200,000 due to smart contracts and blockchain tools.
\item  \textbf{Respondent 5 	}Smart contracts provides an excellent way to our business to records and store business records. We hold millions of client records and despite investing in various security measures in the past, intruders could bypass encryption and firewall making the customers’ records and our business secrets vulnerable to hacking. Smart contracts deploy strong encryption keys that are hard to crack and the entire database comprising client records can be secured through private keys implying that only owners and a few individuals can access the records. Moreover, smart contracts have improved the speed and led to cost-efficiency and offer a better alternative to our legacy systems which we are gradually replacing as blockchain matures. We have been able to eliminate the costly services like using brokers and lawyers since our e-commerce platform sellers can handle all the transactions by themselves with seeking the services or lawyers or third parties who can be costly.  
\item \textbf{Respondent 6} 	As a mortgage company, our business has benefited from smart contract in various way, including more secure, cheap and fast transactions. Today, people can get into their property faster than before since the technology makes the whole process to have minimal hassle. We can digitally agree with the clients to sale even before processing their payment. After the deal is sealed, smart contract easily update properties’ ownership details to show the new owners. The use of a unique key code fair authorization for the original owner make the entire process very secure with minimal conning or fraud cases.  
\item  \textbf{Respondent 7 	}Our insurance business utilizes millions every year in processing and paying out claims. In the past, we have lost missive funds to fraudulent and false claims. Smart contract have come to our recover through supporting the initial policy and enhanced the processing of claims. Our employees and efficiently check for errors in various claims and payouts using the established terms between our company and the claimant. Therefore, it minimizes falling into trap of false claims and errors and drastically makes it cheaper to process claims. With the likelihood of using smart vehicles in future, we anticipate to implement pay-as-you-go strategy to enable immediate activation of claims after our clients incur an accident by promptly processing accident reports, policy information, driving records and licenses to foster rapid payouts to save our clients time and also ensure our company saves time and cost of processing claims and payouts.   
\item  \textbf{Respondent 8 	}Our business processes have greatly benefited from the use of smart contracts. Particularly, stocktaking and our supply chain have been streamlined. We have significantly substantially eliminated in house-thefts since our managers are able to trace any missing products back to the exact place they were located and the time they went missing. Any culpable worker is easy to find and is punished for such activities. We can also be able to view real-time levels of stocks and the period taken to move goods in the supply chain. We use such data to analyze the effectiveness our processes and adjust the stock levels including developing now practices and processes that enhance delivery time. This improves the way we serve our online clients since when order are delivered on time, the clients are happy. We can initiate automatic reordering and easily pay when the orders are received. Ordering and payment information is integrated into our smart contracts to assist us establish the type of products to order in different period of the year.   
\item  \textbf{Respondent 9 	}The use of smart contract in compensating clients for delayed flights is quite aa breakthrough and it resolves challenges associated with manual and traditional bureaucratic processes that are time-consuming and can delay the clients from receiving their money. It enables our clients to book other flights in a fast way since their information is automated and can promptly be compensated. The fear of compensating the wrong client does not arise since the data of our clients is saved duplicated in several databases that cannot be altered by any individual to swindle the clients. Thus, the entire process is more efficient, secure and occurs at outstanding speed.  
\item  \textbf{Respondent 10 	}The use of smart contract in buying and selling of properties such as real estate significant minimizes reduces cases of fraud, deception and conning. Money is held by an escrow which is also aware of terms of contracts established between the buyer and seller. In case the seller does not provide the property based on the descriptions provided in the smart contract. Money will be returned to the buyer. Also, money is not transferred to the seller until when the product has changed ownership. Therefore, smart contracts facilitate secured transactions where both parties are happy and not one swindle the other. Furthermore, since all the paperwork is undertaken digital and is legally binding, smart contract streamline and make the process of buying and selling property efficient and quick to complete.  
\item  \textbf{Respondent 11} 	The major benefit realized from using smart contract in invoice financing is the elimination of intermediaries who are paid and can increase transaction costs. Smart contracts ensure transactions move rapidly while mitigating risks associated with duplication of invoice financing, human errors and fraud. At our company, once the invoice sellers upload invoices and terms, the buyers are able to select the money and the entire transactions are carried out digitally guided by the terms outlined in the smart contracts. Terms cannot be altered once agreed up and transactions are easy to transactions. Information of the buyers and seller of invoice remain undisclosed to any third party which boosts security.   
\end{enumerate}

\subsection{Challenges limiting deployment of Smart Contracts in E-Commerce  }
The participants were asked to offer their opinions of what they believe re the challenges affecting the deployment of smart contracts in e-commerce.  
\textbf{Table 3: Views of the challenges and limitations of smart contracts  }
\begin{enumerate}
\item \textbf{Respondent 1  	}The adoption of smart contracts can be very challenging especially the technical obstacles when migrating and upgrading from legacy systems to blockchain. You need to train your workforce regarding how smart contracts work. Workers are so used to traditional systems and they might resist moving to smart contracts due to fear of learning and understanding new technologies. Besides, smart contracts and blockchain technologies are costly. To effectively deploy the technology, you must invest in talented persons and in the technology itself and before any management or executives invest in the technology, you must convince them the technology will reap the company positive benefits. It is hard to explain to CEOs about benefits smart contract when they cannot figure out what is blockchain in the first place.  
\item \textbf{Respondents 2 	}While smart contracts ensures transactions are streamlined, it can be a major undoing since all the participants need to agree on plainly defined standards. It has been hectic one our clients or our business wants to change the rules or terms of engagement since the changes can only occur after all parties involved agree to the changes. Some parties might take long to respond especially when you are dealing with more than one client within a single smart contract. Also, many of our workers lack sufficient understanding of smart contracts. Our company has been forced to invest money to train them on who smart contracts work. Finding experts to train workers is a challenging processes and once your find one, they charge highly as compared to training on other technologies. Therefore, an organization with insufficient funds should not go for smart contracts, but get prepared since it is an uphill task.
\item \textbf{Respondent 3 	}A find smart contract systems and apps to be incompatible with other existing legacy systems. It is hard for blockchain systems to communicating with our old systems. Their maturity level are questions and most of the apps being developed are full of flaws and need further improvement. Another determining factors for our company to deploy smart contract is performance in validating and processing transactions as well as fraud detection. Smart contracts are a bit slow in handling many transactions without compromising security of data. Therefore, you will be forced to trade off speed or security. In business you need to decide what is important over the other. Furthermore, there are many uncertainties whether contracts enforced through smart contracts are legally binding especially when selling services or goods to third world nations which have not developed blockchain regulation. In case of aa dispute, the only options we have is to utilize dispute resolution due to varied enforceability of smart contracts in different nations.   
\item \textbf{Respondent 4} 	A key question I keep asking myself is about liability and recourse in smart contract. What happens if someone does something wrong with executing the blockchain contract or suffers a loss? Where will they seek a recourse? Smart contract are still developing and immature. They lack a tech-savvy court system to resolve disputes. While some courts in the US and Europe recognize blockchain, particularly cryptocurrency and how smart contract enhance administration of regulation, they lack technical knowledge about the execution of these contracts. However, they need to become aware on how to utilize blockchain evidence when resolving a dispute. I still do not understand how an individual can be made liable. Also, the permanency of smart contract documentation can be hard to alter in case there is a duress or if the contract breaches some regulatory requirements.  
\item \textbf{Respondent 5 	}It is hard to retain all the benefits of blockchain without compromising or trading off others. For instance, blockchain contracts foster transparency among the parties involved since each party will receive a copy of the original document. But it becomes hard to ensure privacy especially when the parties involved do not intend their information to be disclosed.   
\item \textbf{Respondent 6 	}The development of smart contracts remains an opaque process and many people have little knowledge since they need on programing languages like Java, Solidity, Kotlin or Go. Most developers have not released their codes to the public to assist us when creating the contracts. Without source codes, it is challenging for auditors to assess documents and obtain audit trails for various transactions documented by smart contracts. Also, blockchain contracts are perceived to be promising since they generate unpredictable random values and codes.  However, such randomness is exaggerated. I have had cases where miners controlled the generation and release of blocks to benefit themselves. Thus, while smart contract enhance security, sophisticated hackers can still find loops to bypass the security mechanisms utilized.  
\item \textbf{Respondent 7 	}In smart contract, a legal problem can arise when the products or services advertised do not turn out to be as specified in the ad. For instance, when renting out apartments, some tenant might want their money back since the property is not as specified in the advertisement. We have faced a challenge when the landlord refuses to pay back the tenant. Thus, since smart contracts are executed without physical meetings, cases of misrepresentation and fraud can increase. Furthermore, is problematic to find a person to be held responsible for coding errors or when systems malfunction because of unseen challenges.  
\item \textbf{Respondent 8 	}I believe that enforceability is a major challenges that smart contracts are yet to overcome. As blockchain come into existence, they promised decentralization and contracts which do not require permission before being deployed in order to improve the speed of business transactions between various entities located in different geographical locations. The stipulated dispute resolution for contracts vary from one nation to another and can be settled in courts. However, smart contracts do not involve courts but consensus of nodes for the transactions to be agreed upon. A problem emerge regarding how decentralized systems will ensure a consensus about a dispute, especially those residing in different nations. As many of our clients begin to embrace smart contracts and our company utilizes them, we will need to find a way to resolve conflicts involving international transactions. Some nations are rapidly developing legal provisions for blockchain and smart contract while others do not understand these technologies. Furthermore, different interpretations of contract regulation in relation to smart contract continues to present a major challenge.  
\item \textbf{Respondent 9 	}Presently, the maturity level and development of smart contract is somewhat messy and lacks interoperability. Developers use various languages like Python, Java, Solidity and Csharp.  A lot has to be done to achieve standardization of smart contracts. Sometimes you find we have installed or purchase a smart contract product or app which is not compatible with other tools. I believe for smart contract to achieve wide deployments globally, they need to sort out compatibility and standardization challenges by making them to easily get along with other existing systems and emerging technologies.   
\item \textbf{Respondent 10} 	While smart contracts promise superior security mechanisms, this is not the case in some situations. It seems like hackers can still find security loopholes and vulnerabilities that can be exploited. Even though our organization has a robust security strategy, we still fear that our networks can be attacked and result in millions of losses. Besides, we have invested heavily to train our workforce regarding how smart contract works but recent appraisal still reveal that some workers in our company are not comfortable with smart contract. Some even prefer the use of paper-based contracts due to technical complexity of smart contract. Therefore, will feel that our business is losing a lot of funds to blockchain smart contract whereas the returns are low.  
\item \textbf{Respondent 11} 	A major undoing of smart contract is finding the right talent of developers to code the contracts and execute the terms and policy. Hiring people with the right knowledge can be hard since you have to pay them highly. Also, I think the technology lacks regulatory framework for dispute resolution, including international regulation that are standardized across various jurisdictions. Moreover, there is still poor understanding by CEOs and managers about the role played by smart contracts and exactly how they work. Some are hesitant in investing in the technology due to lack of knowledge. Poor understanding is also witnessed among some employees who need thorough training before transitioning fully to the technology.  
\end{enumerate}

\subsection{The future of Smart Contracts and Solutions  }
Indeed, smart contracts have several both from literature examined and the primary data collected. Therefore, it was prudent to seek first-hand opinions from the e-commerce users of this technology regarding what they expect is the future of smart contract and what should be done to enhance their effectiveness.  
\textbf{Table 4: The future and solutions to smart contract challenges  
} 

\begin{enumerate}
\item \textbf{Respondent 1}  	I believe that developers of smart contract technology should work to improve its interoperability to enhance wide-scale adoption and deployment.  Additionally, learning institutions should introduce courses on blockchain and smart contract in order for the future workforce to be well-verse about the technology. In the workplace, businesses should invest in training their workforce about smart contract to minimize resistance during its implementation.  
\item \textbf{Respondents 2 	}Flexibility is a major ball of contention. Smart contracts are increasingly rigidity particularly in altering contract provisions and terms. The technology should allow people to make mistakes and modify any terms that are defined easily without bringing back to traditional hassle and bureaucratic procedures which are lengthy and time-consuming. Furthermore, the cost of hiring training the workforce on smart contract should go down so that many companies can realize return on investment.  
\item \textbf{Respondent 3 	}Experts in the field of smart contract should focus on enhancing the efficiency of the technology to be able to handle several transactions simultaneously in order to same time. Also, security will remain a major determinant in the adoption and retention of blockchain contracts. I read a lot of stuff online regarding what is being done to secure smart contract transactions. These are the most awaited improved to assure the users that their information will not be compromised.  
\item \textbf{Respondent 4} 	I think that legal frameworks and regulations need to be developed to offer guidance in conflict resolution in smart contract. Since the technology is still developing, I expect to see policymakers from different countries coming together to form a body for developing blockchain regulations that can be used across various nations. Also, I expect the see more flexible smart contract were modification or changes can be made easily to the initial terms. Sometimes people make mistakes and smart contracts should allow them to alter some stances where applicable without any deception.  
\item \textbf{Respondent 5} 	My major concern is about privacy and security of data in blockchain contracts. Since the technology fosters decentralization of information, third parties are likely to gain access to sensitive data. Therefore, the developers should seal these loopholes so that the technology can keep its promise of securing transactions.  
\item \textbf{Respondent 6 	}Improvement of knowledge and understanding can significantly improve the use and adoption of smart contract. We need to seem more universities and colleges offering courses on smart contract and blockchain. Also, smart contract tools should be made open source so that individual developers and organizations can improve on the tools and address any bugs. Open source codes can be accessed by auditors for evaluation and tracking of various transactions. Furthermore, the existing security loopholes should be sealed to guarantee safe transactions without exposing customer data.   
\item \textbf{Respondent 7 	}A legal mechanism is needed to ensure that parties involved in a smart contract can seek legal redress irrespective of the country of origin. Sellers need to be assured that they will receive their payment while the buyers can also get their money back in case the product or property agreed upon does not match the indicated specifications. Developers should also ensure that they continuously improve their codes to minimize coding errors experienced in some smart contract and other blockchain applications.  
\item \textbf{Respondent 8} 	More research and development should focus on enhancing security mechanisms of smart contract to secure sensitive information belong to the clients. Also, I expect to see smart contract and other blockchain technologies becoming more efficient and cost-effective. The developers need to find strategies that can bring down the cost of blockchain tools to enable small and medium-sized business deploy them.  
\item \textbf{Respondent 9} 	Further development is needed in the field of blockchain to make the technology mature enough with minimal errors. Developments need to concentrate on security and privacy of information. Also, developers should offer blockchain tools that are compatible with existing systems and can be used in various devices such as laptops, tablets and smartphones.  
\item \textbf{Respondent 10 	}Companies developing blockchain technologies should create trusted systems for e-commerce. The development of trust smart contracts will enhance the economic and social efficacies by fostering financial security as well as enhance the core competency of the systems.  In such systems, data cannot be tampered with. Besides, distributed account ledgers should be offered to resolve cross-business exchanges. Such systems should be standardized to facilitate communication and information sharing between the systems.  
\item \textbf{Respondent 11 	}The global market should invest in talent in blockchain industry for organizations to access competent and skilled workers to push ahead the use of smart contracts. Additionally, regulatory frameworks and guidelines need to be developed to offer laws and guidelines regarding the use of smart contracts. This will foster easier conflict or dispute resolution. Moreover, company managers and CEO need to be aware of market dynamics regarding emerging technologies and their benefits. Leaders should not shy away from learning about smart contracts and its applications in business.  
\end{enumerate}

\section{ DISCUSSION AND CONCLUSION }
Blockchain technologies have recently fueled massive interests in both industry and research due to the ability to enable decentralization of transactions that can be processed minus the need of a trusted intermediary. This paper investigated the applicability of smart contract in e-commerce by obtained data from various industry leaders with understanding about the technology. First the study examined the use cases of deploying smart contracts in e-commerce based on the recruited participants. Findings indicated that smart contracts are deployed to streamline financial transactions including transfer of funds, record-keeping and facilitating financial trade, credit authorization and crowdfunding agreements or Initial Coin Offering (ICOs). Other e-commerce business utilize smart contracts for storing receipts, customer data, managing stock, document property and exchange of ownership and automate payment of rent for electric vehicles. The technology is also deployed for decentralized intelligence, gaming and securing transactions as well as in insurance to encode policy terms for rapid payouts to the client. E-commerce businesses can use smart contracts to manage and secure the buying and selling of products via the internet, shipping of the products and tracking every processes within the supply chain to minimize theft and lost orders. The aviation sector has also found the technology to be important in ensuring processing compensation claims of delayed flights and in real estate, it improves property management. Another emerging use of smart contract include invoice financing.  
This explains why large-scale investments are being made on smart contracts with venture-capitalists funding start-ups to reinvent and transform the conventional business models (Katsiampa, 2019). Ream et al. (2016) explain that the ever-increasing fields of relevance and applications this technology will see many e-commerce businesses customizing the technology to fit their needs. The increased use can be further explained by the benefits delivered by smart contracts. The respondents offered various opinion regarding the benefit of this technology in e-commerce.  
The technology enhances privacy, security and autonomy of transactions be limiting exposure to third parties. Udokwu et al. (2018) explain that blockchain technology allow consumers to purchase products without exposing their personal data. Golosova and Romanovs (2018) attribute the outstanding security to distributed ledge where consumer data is not stored in a centralized server and this minimizes the chances of being hacked or compromised. Also, once 
transactions have been recorded it is impossible to edit, delete or alter the transaction. The permanency of these transactions implies that workers cannot collude to siphon funds and this limits corruption and fraudulent activities since smart contracts offer a perfect audit trail where money can be tracked to the origin. Also, terms of the contract cannot be alter until there is a consensus among the nodes involved. Secure transactions enhances the reputation of businesses making clients satisfied and loyal since their interests are safeguarded. Happy clients improve retention of customers who repeat purchases to foster sound financial performance and profitability.  
Smart contracts were found to improve efficacy, cost-effectiveness and speed of transactions by automating processes while eliminating bureaucratic and lengthy procedures involving third parties that are time-consuming. Cost. Effectiveness is also achieved through reducing the number of workers needed to process transactions and elimination of administration and third party (such as lawyers and brokers) costs. Similar findings are documented by Wang (2016), arguing that the disintermediation process in smart contracts enhance efficiency by addressing critical aspects of transactions ranging from record-keeping, enforcement, auditing and monitoring. Moreover, smart contracts minimise human errors and mistakes since the technology automate most transactions by encoding the terms and conditions that are executed automatically.  
The technology fosters transparency and accountability where transactions are shared among the parties who hold an original copy and buyers can track their products after ordering. Moreover, the storage of transactions in distributed ledger acts as a backup and resilience strategy to foster business continuity and recovery in case of data failures. Similar findings are documented by Giancaspro (2017) who deduce that the absence of central authority is smart contracts and intermediary for validation and collation of transactions coupled with the transparency means that commercial activities undertaken through blockchain contracts in public ledger are visible to all miners. Furthermore, the technology delivers real-time in supply chain to enable managers undertake analytics to obtain knowledge applied of stock management including ordering of stock.  
While smart contract promise to reinvent and transform e-commerce, the technologies face a myriad of challenges and limitations. Golosova and Romanovs (2018) note that the technology is less than a decade old and needs time to mature. There are technical challenges of migrating from legacy systems to blockchain contracts attributed to interoperability challenges coupled with lack of standardisation. Also, some apps and tools are full of bugs and flaws that can expose data 
to hackers. Moreover, Smart contract can be slow and ineffective as compared to conventional legacy systems and networks known for the capability to process thousands of transactions simultaneously like Visa payment system that processes 65,000 transaction per second (Anupam, 2019).  
The immaturity challenge also implies that Blockchain technology lacks sufficient legal frameworks and regulatory mechanism, especially for international deals. It remains uncertain whether smart contracts can be enforced by law legally bindings. Also, it is not clear whether unsatisfied party can seek legal redress in a court of law and who should be held liable since there are no tech-savvy courts across the globe. Moreover, different interpretations of contract regulations in relation to smart contract continues to present a major challenge. According to Meijer (2020), there is insufficient clarity of the regulatory frameworks and laws governing blockchain technologies and this affects mass adoption in business. Nonetheless, Golosova and Romanovs (2018) have a different opinion indicating that digitalization and development of novel technologies are occurring in a heavily environment due to the long-standing regulations enacted by various regimes intended to safeguard the consumers as well as property rights. Salmerón-Manzano and Manzano-Agugliaro (2019) point out that smart contracts need to tackle various issues like jurisdictional, trial and enforcement risks. 
Additionally, workers, managers and some CEOs lack sufficient understanding about technology and raise doubts whether smart contracts will benefit businesses. Therefore, there is hesitation in investing in the blockchain contracts. Besides, it is costly to hire and training workers on smart contract due to the immaturity of the technology and scarcity of talent in the arena. Meijer (2020) found that the demand for blockchain-related talent had skyrocketed by about 2000 percent  from 2017 to 2020. Besides, smart contracts are still considered to be opaque to many and this is escalated by absence of original codes to enable robust auditing. Furthermore, while permanency of blockchain records is a major benefit in evading corruption and fraud, it is hectic for parties to alter the terms and provisions after they agree to modify something after discovering an error.  
Privacy issue comes into play for decentralised ledger transactions where information is likely to leak to other third parties. Also, there are security loopholes that can be exploited by sophisticated hackers to generate their order blocks for fraudulent activities. Conoscenti et al. (2016) explains that only limited scenarios of blockchain contracts offer adequate security mechanism to cope emerging security threats. While blockchain technologies are more secure as 
compared to the conventional systems, sophisticated hackers are still breaching businesses, applications and systems deploying blockchain technologies (Anupam, 2019). Evidently, while smart contract provides outstanding solutions to e-commerce, policymakers, researchers, practitioners and other stakeholders, need to invest effort and time to foster growth to the technologies and resolve existing challenge to enable global adoption of smart contract by small and big businesses.

\textbf{5.1	Recommendations  }

Notwithstanding the rapid developments of blockchain contracts and their beneficial impacts, there are several challenges that need to be addressed. The developers of smart contracts and blockchain need to collectively work in enhance interoperability and security loopholes to encourage widespread of the adoption of these tools in e-commerce and other sectors. The detection of security loopholes and sealing them will required extensive software engineering and data analytics. Smart contracts tools and systems should be able to run in different environments and devices to ensure compatibility with existing systems.  
Besides, more research should focus on enhancing efficiency of smart contract by developing tools that reduce the cost of the technology, energy consumption and storage.  Researchers should follow open source model where codes are available to the public for enhancement and use.  Also, it is vital that higher learning institutions begin offering robust courses on blockchain to ensure the labour market has sufficient workers with understanding on blockchain technologies. Within the workplace, HR and managers should work together to offer training to their workforce to foster smooth transition into smart contracts.  
The legal environment must work fast to establish policies, standards and regulation for smart contracts that can be deployed internationally, since the majority of transactions involve geographically dispersed parties. Courts and policymakers will need to establish the pathways utilized to introduce, validate and admit evidences regarding smart contracts. Moreover, smart contract regulations need be enforced to enable regulators access private information in order to track underlying transactions. Also, courts and lawyers need to become tech-savvy in relation to smart contracts. Personnel should receive training on various modalities of smart contracts and how they can obtain pieces of evidence to resolve various conflicts.

\section{References }

\begin{enumerate}
\item Abadi, M., Chu, A., Goodfellow, I., McMahan, H.B., Mironov, I., Talwar, K. and Zhang, L., 2016, October. Deep learning with differential privacy. In Proceedings of the 2016 ACM SIGSAC Conference on Computer and Communications Security (pp. 308-318). 
\item Adu, P. 2019. A step-by-step guide to qualitative data coding. Abingdon, Oxon; New York, NY : Routledge.  
\item Ali, O., Ally, M. and Dwivedi, Y., 2020. The state of play of blockchain technology in the financial services sector: A systematic literature review. International Journal of Information Management, 54, p.102199. 
\item Anupam, S. 2019. What Are The Major Limitations, Challenges In Blockchain? Retrieved from INC42: https://inc42.com/features/what-are-the-major-limitations-challenges-in-blockchain/ 
\item Atzei, N., Bartoletti, M. and Cimoli, T., 2017, April. A survey of attacks on ethereum smart contracts (sok). In International conference on principles of security and trust (pp. 164-186). Springer, Berlin, Heidelberg. 
\item BarNir, A., Gallaugher, J. M., and Auger, P. 2003. Business process digitization, strategy, and the impact of firm age and size: the case of the magazine publishing industry. Journal of Business Venturing, 18(6), 789-814. 
\item Beneki, C., Koulis, A., Kyriazis, N.A. and Papadamou, S., 2019. Investigating volatility transmission and hedging properties between Bitcoin and Ethereum. Research in International Business and Finance, 48, pp.219-227. 
\item Burnard, P., Gill, P., Stewart, K., Treasure, E. and Chadwick, B., 2008. Analysing and presenting qualitative data. British dental journal, 204(8), pp.429-432. 
\item Carson, B., Romanelli, G., Walsh, P. and Zhumaev, A., 2018. Blockchain beyond the hype: What is the strategic business value. McKinsey and Company, pp.1-13. 
\item Carson, B., Romanelli, G., Walsh, P., and Zhumaev, A. 2018. Blockchain beyond the hype: What is the strategic business value? Retrieved from Mckinsey: https://www.mckinsey.com/business-functions/mckinsey-digital/our-insights/blockchain-beyond-the-hype-what-is-the-strategic-business-value
\item Casino, F., Dasaklis, T.K. and Patsakis, C., 2019. A systematic literature review of blockchain-based applications: current status, classification and open issues. Telematics and Informatics, 36, pp.55-81. 
\item Chepurnoy, A. and Saxena, A., 2019. Multi-stage Contracts in the UTXO Model. In Data Privacy Management, Cryptocurrencies and Blockchain Technology (pp. 244-254). Springer, Cham. 
\item Cole, R., Stevenson, M. and Aitken, J., 2019. Blockchain technology: implications for operations and supply chain management. Supply Chain Management: An International Journal, 24(4), pp.469-483. 
\item Conoscenti, M., Vetro, A. and De Martin, J.C., 2016, November. Blockchain for the Internet of Things: A systematic literature review. In 2016 IEEE/ACS 13th International Conference of Computer Systems and Applications (AICCSA) (pp. 1-6). IEEE. 
\item Creswell, J. W. 2007. Qualitative inquiry and research design. London: Sage Publications. 
\item Creswell, J. W. 2014. Research design: Qualitative, quantitative, and mixed methods approaches. Thousand Oaks: SAGE. 
\item Daniel, E., and Wilson, H. 2002. Adoption intentions and benefits realised: a study of e‐commerce in UK SMEs. Journal of Small Business and Enterprise Development. 
\item De Vries, A., 2019. Renewable energy will not solve bitcoin’s sustainability problem. Joule, 3(4), pp.893-898. 
\item Edelman, R. 2017. 2017 Edelman trust barometer: Global report. 
\item Fries, M., and Paal, B. P. 2019. Smart contracts. Mohr Siebeck. 
\item Gatteschi, V., Lamberti, F., Demartini, C., Pranteda, C. and Santamaría, V., 2018. Blockchain and smart contracts for insurance: Is the technology mature enough?. Future Internet, 10(2), p.20. 
\item Giancaspro, M., 2017. Is a ‘smart contract’really a smart idea? Insights from a legal perspective. Computer law and security review, 33(6), pp.825-835. 
\item Golosova, J. and Romanovs, A., 2018, November. The advantages and disadvantages of the blockchain technology. In 2018 IEEE 6th workshop on advances in information, electronic and electrical engineering (AIEEE) (pp. 1-6). IEEE. 
\item Iansiti, M., and Lakhani, K. R. 2017. The Truth About Blockchain. Retrieved from Havard Business Review: https://hbr.org/2017/01/the-truth-about-blockchain 
\item Ismanto, L., Ar, H. S., Fajar, A. N., and Bachtiar, S. 2019. Blockchain as E-Commerce Platform in Indonesia. In Journal of Physics: Conference Series (Vol. 1179, No. 1, p. 012114). IOP Publishing. 
\item Katsiampa, P., 2019. An empirical investigation of volatility dynamics in the cryptocurrency market. Research in International Business and Finance, 50, pp.322-335. 
\item Khezr, S., Moniruzzaman, M., Yassine, A. and Benlamri, R., 2019. Blockchain technology in healthcare: A comprehensive review and directions for future research. Applied sciences, 9(9), p.1736. 
\item King, N., Horrocks, C., and SAGE Publishing. 2010. Interviews in qualitative research. London [i pozostałe: SAGE. 
\item Kouhizadeh, M., Saberi, S. and Sarkis, J., 2020. Blockchain technology and the sustainable supply chain: Theoretically exploring adoption barriers. International Journal of Production Economics, 231, p.107831. 
\item Meijer, C. R. 2020. Remaining challenges of blockchain adoption and possible solutions. Retrieved from Finextra: https://www.finextra.com/blogposting/18496/remaining-challenges-of-blockchain-adoption-and-possible-solutions 
\item Mengelkamp, E., Notheisen, B., Beer, C., Dauer, D. and Weinhardt, C., 2018. A blockchain-based smart grid: towards sustainable local energy markets. Computer Science-Research and Development, 33(1-2), pp.207-214. 
\item Nikolić, I., Kolluri, A., Sergey, I., Saxena, P. and Hobor, A., 2018, December. Finding the greedy, prodigal, and suicidal contracts at scale. In Proceedings of the 34th Annual Computer Security Applications Conference (pp. 653-663). 
\item Notland, J. S., Notland, J. S., and Morrison, D. 2020. The Minimum Hybrid Contract (MHC) Combining Legal and Blockchain Smart Contracts. In Proceedings of the Evaluation and Assessment in Software Engineering (pp. 390-397). 
\item Pan, X., Pan, X., Song, M., Ai, B. and Ming, Y., 2020. Blockchain technology and enterprise operational capabilities: An empirical test. International Journal of Information Management, 52, p.101946. 
\item Ream, J., Chu, Y. and Schatsky, D., 2016. Upgrading blockchains: Smart contract use cases in industry. Retrieved December, 12, p.2017. 
\item Ryan, P.A., 2017. Smart contract relations in e-commerce: legal implications of exchanges conducted on the blockchain. Technology Innovation Management Review. 
\item Salmerón-Manzano, E. and Manzano-Agugliaro, F., 2019. The role of smart contracts in sustainability: worldwide research trends. Sustainability, 11(11), p.3049. 
\item Savelyev, A. 2017. Contract law 2.0:‘Smart’contracts as the beginning of the end of classic contract law. Information  and  Communications Technology Law, 26(2), 116-134. 
\item Singh, A., Parizi, R. M., Zhang, Q., Choo, K. K. R., and Dehghantanha, A. 2020. Blockchain smart contracts formalization: Approaches and challenges to address vulnerabilities. Computers  and  Security, 88, 101654. 
\item Suliman, A., Husain, Z., Abououf, M., Alblooshi, M. and Salah, K., 2018. Monetization of IoT data using smart contracts. IET Networks, 8(1), pp.32-37. 
\item Szabo, N. 1997. Formalizing and securing relationships on public networks. First Monday. 
\item Tanja, M. 2018. How Can Blockchain Smart Contracts Be Used in e-Commerce? Retrieved from Stockholm: https://stockholm.bc.events/en/news/how-can-blockchain-smart-contracts-be-used-in-e-commerce-91278 
\item Tapscott, D. and Tapscott, A., 2016. Blockchain revolution: how the technology behind bitcoin is changing money, business, and the world. Penguin. 
\item Tapscott, D. and Tapscott, A., 2017. How blockchain will change organizations. MIT Sloan Management Review, 58(2), p.10. 
\item Tapscott, D., and Tapscott, A. 2016. Blockchain revolution: how the technology behind bitcoin is changing money, business, and the world. Penguin. 
\item Tönnissen, S. and Teuteberg, F., 2020. Analysing the impact of blockchain-technology for operations and supply chain management: An explanatory model drawn from multiple case studies. International Journal of Information Management, 52, p.101953. 
\item Udokwu, C., Kormiltsyn, A., Thangalimodzi, K. and Norta, A., 2018. An exploration of blockchain enabled smart-contracts application in the enterprise. Technical Report. 
\item Walch, A., 2015. The bitcoin blockchain as financial market infrastructure: A consideration of operational risk. NYUJ Legis.  and  Pub. Pol'y, 18, p.837. 
\item Wang, B., 2016. Blockchain and the law. Internet Law Bulletin, 19(1), pp.250-254. 
\item   Wang, S., Yuan, Y., Wang, X., Li, J., Qin, R. and Wang, F.Y., 2018, June. An overview of smart contract: architecture, applications, and future trends. In 2018 IEEE Intelligent Vehicles Symposium (IV) (pp. 108-113). IEEE. 
\item Yaga, D., Mell, P., Roby, N. and Scarfone, K., 2019. Blockchain technology overview. arXiv preprint arXiv:1906.11078. 
\item Zhang, Y., Bian, J., and Zhu, W. 2013. Trust fraud: A crucial challenge for China’s e-commerce market. Electronic Commerce Research and Applications, 12(5), 299-308. 
\end{enumerate}

\section{APPENDIX }
\subsection{Interview Guide  }
\begin{enumerate}
\item Does your business use smart contracts? 
\item Explain the areas in your business where smart contracts are applied. 
\item What are the benefits and reasons for using smart contracts in e-commerce? 
\item How has smart contract improve business processes and efficiency? 
\item What are the major hindrance limiting the use of smart contract in e-commerce? 
\item Is the technology mature enough? 
\item Comment on the legal framework on smart contracts and limitations? 
\item What do you believe can be done to resolve the challenges you have discussed? 
\item What is the future of smart contracts? 
\item What development could you like to see in smart contract field? 
\end{enumerate}

\end{document}